\documentclass{JHEP3}

\usepackage{epsfig} 
\usepackage{cite}

\newcommand{\beq}{\begin{equation}}
\newcommand{\eeq}{\end{equation}}
\newcommand{\bea}{\begin{eqnarray}}
\newcommand{\eea}{\end{eqnarray}}
\newcommand{\nn}{\nonumber}

\def\eqn#1{Eq.~(\ref{#1})}
\def\eqns#1#2{Eqs.~(\ref{#1}) and~(\ref{#2})}

\def\fig#1{Fig.~{\ref{#1}}}
\def\sec#1{Sect.~{\ref{#1}}}

\def\bom#1{{\mbox{\boldmath $#1$}}}

\def\spa#1.#2{\left\langle#1\,#2\right\rangle}
\def\spb#1.#2{\left[#1\,#2\right]}

\def\tr{\mathop{\rm tr}\nolimits}
\def\tree{{\rm tree}}
\def\oneloop{{1 \mbox{-} \rm loop}}
\def\ct{{\rm ct}}
\def\re{{\rm Re}}
\def\eps{\epsilon}
\def\ord{{\cal O} }
\def\cA{{\cal A}}
\def\cM{{\cal M}}
\def\cP{{\cal P}}
\def\d{{\rm d}}
\def\i{{\rm i}}
\def\ib{{\bar i}}
\def\qb{{\bar q}}
\def\Qb{{\bar Q}}

\def\nn{\nonumber}
\def\mathswitchr#1{\relax\ifmmode{\mathrm{#1}}\else$\mathrm{#1}$\fi}

\renewcommand{\perp}{{\rm T}}
\newcommand\sss{\scriptscriptstyle}
\newcommand\as{\alpha_{\sss S}}
\newcommand\aem{\alpha_{\rm em}}
\newcommand\gs{g_{\sss S}}
\newcommand\mh{m_{\sss H}}
\newcommand\mgg{m_{\sss \gamma\gamma}}
\newcommand\et{E_\perp}
\newcommand\pt{p_\perp}

\newcommand\ptgg{p_{\gamma\gamma\perp}}
\newcommand\ptjet{p_{{\rm jet}\perp}}
\newcommand\jet{+ {\rm jet}}

\title{QCD radiative corrections to prompt diphoton production in
association with a jet at hadron colliders}

\author{Vittorio Del Duca\\
Istituto Nazionale di Fisica Nucleare, Sez. di Torino\\
via P. Giuria, 1 - 10125 Torino, Italy\\
        E-mail: \email{delduca@to.infn.it}}

\author{Fabio Maltoni\thanks{Mail Address: Dipartimento di Fisica,
Terza Universit\`a di Roma, via della Vasca Navale, 84 - 00146
Rome, Italy.}\\
Centro Studi e Ricerche ``Enrico Fermi'' \\
via Panisperna, 89/A - 00184 Rome, Italy \\
E-mail: \email{maltoni@fis.uniroma3.it}}

\author{Zolt\'an Nagy\\
Institute of Theoretical Science\\
5203 University of Oregon\\
Eugene, OR  97403-5203, USA\\
E-mail: \email{nagyz@physics.uoregon.edu}}

\author{Zolt\'an Tr\'ocs\'anyi\thanks{Sz\'echenyi fellow of the Hungarian
Ministry of Education}\\ University of Debrecen and\\
Institute of Nuclear Research of the Hungarian Academy of Sciences\\
H-4001 Debrecen, PO Box 51, Hungary\\
        E-mail: \email{zoltan@zorro.atomki.hu}}

\abstract{We compute the next-to-leading order corrections in $\as$ to
prompt diphoton production in association with a jet at hadron colliders.
We use a next-to-leading order general-purpose partonic Monte Carlo 
event generator that allows the computation of a rate differential 
in the produced photons and hadrons.}

\preprint{{~DFTT 43/2002}\\{~RM3-TH/03-1}\\{~hep-ph/0303012}}

\begin{document}

\section{Introduction}
\label{sec:intr}

Higgs production in association with a jet of high transverse energy
$\et$ with a subsequent decay into two photons, $pp\to H\jet \to
\gamma\gamma\jet$, is considered a very promising discovery channel for
a Higgs boson of intermediate mass (100\,GeV $\lesssim \mh \lesssim$ 
140\,GeV)~\cite{Dubinin:1997rq,Abdullin:1998er}. In fact if a
high-$\pt$ jet is present in the final state, the photons are
more energetic than in the inclusive channel and the
reconstruction of the jet allows for a more precise determination of
the interaction vertex.  Moreover, the presence of the jet offers the
advantage of being more flexible with respect to choosing suitable
acceptance cuts to curb the background.  These advantages compensate
for the loss in the production rate. The analysis presented in
Refs.~\cite{Dubinin:1997rq,Abdullin:1998er} was done using
leading-order perturbative predictions for both the
signal~\cite{Ellis:1988xu} and for the background~\cite{Boos:1994xb},
although anticipating large radiative corrections which were taken into
account by using a constant $K^{\rm NLO} = 1.6$ factor for both the
signal and the background processes.

Since the first analysis was proposed, the radiative corrections to the
signal have been computed \cite{deFlorian:1999zd}. However,
the irreducible $p p\to\gamma\gamma\jet$ background is still known
only at leading order, where it proceeds via the sub-processes
$q \bar q \to \gamma\gamma g$ and $qg\to \gamma\gamma q$ and is
dominated by the latter that benefits from the large gluon luminosity.
The quark-loop mediated $gg\to g \gamma\gamma$ sub-process, which is
$\ord(\as^3)$ and thus formally belongs to the next-to-next-to-leading order
(NNLO) corrections, might have been
significant due to the large gluon luminosity. It has been computed and
found to yield a modest contribution~\cite{deFlorian:1999tp,Balazs:1999yf}.  
Thus it is not unreasonable to expect that a calculation of the unknown 
next-to-leading order (NLO) corrections 
provide a reliable quantitative estimate of the irreducible
$pp\to\gamma\gamma\jet$ background. In this paper we present the
NLO corrections to $pp\to\gamma\gamma\jet$. 
Although this background will eventually be measured at the Large Hadron
Collider, it is still important to have a better understood theoretical
prediction in order to optimize the various detector-dependent
selection and isolation parameters used for Higgs searches.

The necessary ingredients for the computations presented here have been
known in the literature for some time. At NLO accuracy we must consider
the tree-level subprocesses with two partons and two photons in the
final state, namely $q\qb\to g g \gamma\gamma$ and $q\qb\to Q \Qb
\gamma\gamma$ together with all the crossing-related subprocesses.
The corresponding contributions are termed real corrections.
We express the matrix elements in terms of amplitudes of fixed
helicities, compute the square and the sum over helicities
numerically. For maximally-helicity-violating
configurations, the amplitudes were first derived in
Ref.~\cite{Parke:1986gb}. The remaining helicity configurations were
computed in Ref.~\cite{Barger:1990yd}. In our calculations we used the
amplitudes presented in Ref.~\cite{DelDuca:2000pa}, where
all helicity amplitudes for processes involving $n$ photons and $6-n$
partons ($n=1,2,3,4$) were re-calculated using the notation of
Ref.~\cite{Mangano:1991by}.

At NLO we also need the interference between one-loop and tree
contributions to the Born-level subprocesses $q\qb\to g \gamma\gamma$
and $qg\to \gamma\gamma q$, which constitute the virtual corrections.
The one-loop helicity amplitudes are decomposed in terms of primitive
amplitudes for the two-quark three-gluon one-loop
amplitudes~\cite{Bern:1995fz,Signer:1995nk}. The explicit form of the
decomposition was presented in Ref.~\cite{DelDuca:2000pa}. 

In a NLO computation both the real and the virtual corrections are
separately divergent in four dimensions, but their sum is finite for
infrared-safe observables. There are several methods to compute this
finite correction. We used the dipole subtraction
method~\cite{Catani:1996vz} slightly modified for better numerical
control~\cite{Nagy:1998bb} as implemented in the {\tt NLOJET++}
package~\cite{Nagy:2001fj}.

In \sec{sec:xsect} the cross sections for $pp\to\gamma\gamma\jet$ 
production at leading order and at NLO are described;
in \sec{sec:results} several distributions of eventual phenomenological
interest, as well as the behaviour of the cross section under
renormalization and factorization scale variations, are analysed, with and 
without radiative corrections;
in \sec{sec:conclusions} we present our conclusions.

\section{Cross section formul\ae}
\label{sec:xsect}

In this section we spell out the explicit cross section formul\ae\ that
are used in our computation. First we give the leading order production
rate, then the rates used in the calculation of the radiative
corrections: the real and virtual contributions and and certain
Born-level matrix element expressions needed for the construction of
the dipole subtraction terms.
 
\subsection{The leading order production rate}
\label{sec:lorate}

Here we summarize what is necessary to compute $pp\to\gamma\gamma\jet$
at leading order. All the relevant parton subprocesses can be obtained,
directly or through crossing symmetry, from the tree-level scattering
amplitude for $q\qb\to\gamma\gamma g$. The colour decomposition of amplitudes 
involving two quarks, two photons and a gluon is~\cite{Mangano:1991by}
\beq
{\cal A}_5^\tree(1_\qb,2_q;3_g;4_\gamma,5_\gamma) =
2 \gs e^2 Q_q^2 \, (T^{a_3})_{i_2}^{\ib_1}\,
A_5^\tree(1_\qb,2_q;3_g;4_\gamma,5_\gamma)\:,
\label{TwoQuarkGluonPhot}
\eeq
where $eQ_q$ is the electromagnetic charge of the quark $q$, $\gs$ is
the strong coupling and $T^a$ are the generators of the $SU(N_c)$
algebra in the fundamental representation\footnote{We normalise the
fundamental representation matrices by taking 
$\tr(T^aT^b) = T_R \delta^{ab}$, with $T_R = 1$. For the $C_i$ quadratic
Casimir operators we have $C_F= T_R(N_c^2-1)/N_c= (N_c^2-1)/N_c$ in
the fundamental and $C_A=2\,T_R N_c=2\,N_c$ in the adjoint representation.}. 
On the right-hand side of
\eqn{TwoQuarkGluonPhot} $A_n^\tree$ is the colour-stripped sub-amplitude
which we give for fixed helicities of the external particles, with
all particles taken as outgoing. Configurations with all of the
external particles of equal helicity, or all but one, do not contribute
at tree level.  For maximally-helicity-violating configurations,
$(-,-,+,\dots,+)$, the sub-amplitudes
are~\cite{Parke:1986gb,Mangano:1991by,wu}
\beq
A_5^\tree(1_\qb^+,2_q^-;3_g;4_\gamma,5_\gamma)
= \i{\spa1.i \spa2.i^3 \over \spa1.2\spa2.3 \spa3.1}
{\spa2.1 \over \spa2.4\spa4.1} {\spa2.1 \over \spa2.5\spa5.1}
\label{mhvphotamp}
\eeq
where we make explicit only the helicity of the quark line, and where 
$i$ is the gluon or photon of negative helicity.
\eqn{mhvphotamp} defines the only independent sub-amplitude, 
all of the other sub-amplitudes are obtained by relabeling and
by use of the discrete symmetries of parity inversion and charge
conjugation.  Parity inversion flips the helicities of all particles.
It is accomplished by the ``complex conjugation'' operation (indicated
with a superscript ${}^\dagger$) defined such that $\spa{i}.j
\leftrightarrow \spb{j}.i$, but explicit factors of $\i$ are not
conjugated to $-\i$. A factor of $-1$ must be included
for each pair of quarks participating in the amplitude.  Charge
conjugation changes quarks into anti-quarks without inverting
helicities.  In addition, there is a reflection symmetry in the colour
ordering such that
\beq
A_5^\tree(1_\qb,2_q;3_g;4_\gamma,5_\gamma) =
-A_5^\tree(2_q,1_\qb;3_g;4_\gamma,5_\gamma)\:.
\label{refl}
\eeq
Using \eqns{TwoQuarkGluonPhot}{mhvphotamp}, the squared amplitude for
$q\qb\to\gamma\gamma g$, summed over colours and helicities, is
\bea
&&
\sum_{\rm col,\,hel} |\cA_5^\tree(1_\qb,2_q;3_g;4_\gamma,5_\gamma)|^2 =
8 e^4 Q_q^4 \gs^2 N_c C_F {1\over s_{12} s_{23} s_{13}}
{s_{12}\over s_{14} s_{24}} {s_{12}\over s_{15} s_{25}} 
\nn\\ &&\qquad\qquad
\times \Big[ s_{13} s_{23} (s_{13}^2 + s_{23}^2) +
s_{14} s_{24} (s_{14}^2 + s_{24}^2) + s_{15} s_{25} (s_{15}^2 + s_{25}^2)
\Big]\:.
\label{squamp}
\eea
In order to obtain the production rate for $q\qb\to\gamma\gamma g$, we
cross to the physical region and take 1 as the incoming quark
and 2 as the incoming antiquark,
\beq
\d\sigma(q\qb) =
{1\over 2\hat{s}}\ \d\cP_3\ {1\over 2}\ {1\over 4N_c^2}\
\sum_{\rm col,\,hel} |\cA_5^\tree(1_\qb,2_q;3_g;4_\gamma,5_\gamma)|^2\:,
\label{prodrate}
\eeq
with $\hat s = (p_1+p_2)^2$, and where we make explicit the average
over initial colours and helicities, and the symmetry factor for the
final-state photons. $\d\cP_3$ is the 3-particle phase space for the 
final state,
\beq
\d{\cal P}_3 = \prod_{i=3}^5 {\d^3 p_i\over (2\pi)^3 2E_i}\,
(2\pi)^4 \, \delta^4(p_1 + p_2 - p_3 - p_4 - p_5)\:,
\label{3phase}
\eeq
that we generate uniformly.

The production rates for the subprocesses $q g\to q \gamma\gamma$ and
$\qb g\to \qb \gamma\gamma$ are obtained from \eqn{prodrate} by a
suitable relabelling ($1 \leftrightarrow 3$ and $2 \leftrightarrow 3$,
respectively) of the entries in the squared amplitude (\ref{squamp}),
and by replacing the colour average $N_c^2$ with $N_c(N_c^2-1)$.  

\subsection{The NLO production rate}
\label{sec:nlorate}

At NLO we must consider the tree-level subprocesses with two partons
and two photons in the final state, namely $q\qb\to g g \gamma\gamma$
and $q\qb\to Q \Qb \gamma\gamma$ and all of the crossing-related
subprocesses, and the interference between one-loop and tree
contributions to $q\qb\to g \gamma\gamma$, and crossing-related
subprocesses. Let us first consider the subprocesses with four
final-state particles.


For $q\qb\to g g \gamma\gamma$, the colour decomposition is 
\bea
&&
{\cal A}_6^\tree(1_\qb,2_q;3_g,4_g;5_\gamma,6_\gamma) =
2 \gs^2 e^2 Q_q^2 
\nn \\ &&\qquad
\times
\Big[
(T^{a_3} T^{a_4})_{i_2}^{\ib_1}\,
A_6^\tree(1_\qb,2_q;3_g,4_g;5_\gamma,6_\gamma)
+ (T^{a_4} T^{a_3})_{i_2}^{\ib_1}\,
A_6^\tree(1_\qb,2_q;4_g,3_g;5_\gamma,6_\gamma)
\Big]\:.
\hspace*{2em}
\label{TwoQuarkTwoGluonPhot}
\eea
The sub-amplitudes for the maximally-helicity-violating configurations
$(-,-,+,+,+,+)$ have similar forms as
in~\eqn{mhvphotamp}~\cite{Parke:1986gb,Mangano:1991by,wu}.
Those for the configurations $(-,-,-,+,+,+)$ are more complicated,
and have been computed in Refs.~\cite{Barger:1990yd,DelDuca:2000pa}. For
each helicity configuration, the squared amplitude for
$q\qb\to g g \gamma\gamma$, summed over colours, is
\bea
&&
\sum_{\rm col} |\cA_6^\tree(1_\qb,2_q;3_g,4_g;5_\gamma,6_\gamma)|^2 =
4 e^4 Q_q^4 \gs^4\ C_F
\nn\\ && \qquad
\times \Big\{ N_c C_F
\left[|A_6^\tree(1_\qb,2_q;3_g,4_g;5_\gamma,6_\gamma)|^2
+ |A_6^\tree(1_\qb,2_q;4_g,3_g;5_\gamma,6_\gamma)|^2 \right]
\nn\\ &&\qquad
- 2 \re \left[A_6^\tree(1_\qb,2_q;3_g,4_g;5_\gamma,6_\gamma)^*
A_6^\tree(1_\qb,2_q;4_g,3_g;5_\gamma,6_\gamma)\right] \Big\}\:. 
\label{squamp6}
\eea
As in \sec{sec:lorate}, in order to obtain the production rate for
$q\qb\to\gamma\gamma g g$ we cross to the physical region and take 1 as
the incoming quark and 2 as the incoming antiquark,
\beq
\d\sigma(q\qb\to\gamma\gamma g g) =
{1\over 2\hat{s}}\ \d\cP_4\ {1\over 4}\ 
{1\over 4N_c^2}\ \sum_{\rm col,\,hel} 
|\cA_6^\tree(1_\qb,2_q;3_g,4_g;5_\gamma,6_\gamma)|^2\:,
\label{6prodrate}
\eeq
where we made explicit the symmetry factor for the final-state photons 
and gluons. $\d\cP_4$ is the 4-particle phase space for the 
final state,
\beq
\d{\cal P}_4 = \prod_{i=3}^6 {\d^3 p_i\over (2\pi)^3 2E_i}\,
(2\pi)^4 \, \delta^4(p_1 + p_2 - p_3 - p_4 - p_5 - p_6)\:,
\label{4phase}
\eeq
which we generate from the three-parton phase space \eqn{3phase} with a
subsequent dipole splitting~\cite{Catani:1996vz}.
The production rates for the subprocesses $q g\to q g \gamma\gamma$ and
$\qb g\to \qb g \gamma\gamma$ and $g g\to q \qb \gamma\gamma$
are obtained from \eqn{6prodrate} by a suitable relabelling of the 
entries in the squared amplitude (\ref{squamp6}), by changing the
symmetry factor to that of photons only, and by a suitable replacement
of the colour average.

The colour decomposition for the subprocesses with two pairs of quarks
of different flavour, $q\qb\to Q \Qb \gamma\gamma$, is
\beq
\cA_6^\tree(1_\qb,2_q;3_\Qb,4_Q;5_\gamma,6_\gamma) = 2 e^2 \gs^2
\,(T^a)_{i_2}^{\ib_1}\,(T^a)_{i_4}^{\ib_3} \,
A_6^\tree(1_\qb,2_q;3_\Qb,4_Q;5_\gamma,6_\gamma)\:.
\label{FourQuark}
\eeq
Note that in \eqn{FourQuark} the electromagnetic charges do not factor
out, and thus are kept within the sub-amplitudes.  The sub-amplitudes
for the maximally-helicity-violating configurations $(-,-,+,+,+,+)$ are
listed in Ref.~\cite{Mangano:1991by,DelDuca:2000pa}. 
Those for the configurations $(-,-,-,+,+,+)$ can be found in
Ref.~\cite{Barger:1990yd,DelDuca:2000pa}. For each
helicity configuration, the squared amplitude for 
$q\qb\to Q \Qb \gamma\gamma$, summed over colours, is
\beq
\sum_{\rm col} |\cA_6^\tree(1_\qb,2_q;3_\Qb,4_Q;5_\gamma,6_\gamma)|^2
= 4 e^4 \gs^4 N_c C_F |A_6^\tree(1_\qb,2_q;3_\Qb,4_Q;5_\gamma,6_\gamma)|^2\:.
\label{squamp4quark}
\eeq
The production rate for $q\qb\to Q \Qb \gamma\gamma$ has the same form as
\eqn{6prodrate}, up to the symmetry factor which is for photons only,
\beq
\d\sigma(q\qb\to Q \Qb \gamma\gamma) =
{1\over 2\hat{s}}\ \d\cP_4\ {1\over 2}\ 
{1\over 4N_c^2}\ \sum_{\rm col,\,hel} 
|\cA_6^\tree(1_\qb,2_q;3_\Qb,4_Q;5_\gamma,6_\gamma)|^2\:.
\label{4quarkrate}
\eeq
The production rates for $q Q\to q Q \gamma\gamma$, and for the
substitution of one or of both of the incoming quarks with antiquarks,
are obtained by suitable relabelings of the entries in 
\eqns{squamp4quark}{4quarkrate}.

For two quark pairs of equal flavour, we must antisymmetrise 
\eqn{FourQuark} by subtracting the same expression with the colour and 
momentum labels of the quarks exchanged,
\bea
\cA_6^\tree(1_\qb,2_q;3_\qb,4_q;5_\gamma,6_\gamma) &=& 2 e^2 \gs^2
\Bigg[
(T^a)_{i_2}^{\ib_1}\,(T^a)_{i_4}^{\ib_3} \,
A_6^\tree(1_\qb,2_q;3_\qb,4_q;5_\gamma,6_\gamma) 
\nn \\ && \qquad
- (T^a)_{i_4}^{\ib_1}\,(T^a)_{i_2}^{\ib_3} \,
A_6^\tree(1_\qb,4_q;3_\qb,2_q;5_\gamma,6_\gamma) \Bigg] \:. 
\label{FouridQuark}
\eea
Then for each helicity configuration the squared amplitude for 
$q\qb\to q \qb \gamma\gamma$, summed over colours, is
\bea
&&
\sum_{\rm col} |\cA_6^\tree(1_\qb,2_q;3_\qb,4_q;5_\gamma,6_\gamma)|^2
= 4 e^4 \gs^4 (N_c^2-1) 
\nn \\ &&\qquad
\times \Bigg\{
  |A_6^\tree(1_\qb,2_q;3_\qb,4_q;5_\gamma,6_\gamma)|^2
+ |A_6^\tree(1_\qb,4_q;3_\qb,2_q;5_\gamma,6_\gamma)|^2
\nn \\ &&\qquad\quad
+ \delta_{h_2 h_4} {2\over N_c} \re
\left[ A_6^\tree(1_\qb,2_q;3_\qb,4_q;5_\gamma,6_\gamma)^*
       A_6^\tree(1_\qb,4_q;3_\qb,2_q;5_\gamma,6_\gamma) \right] \Bigg\}\:,
\label{squamp4idquark}
\eea
where the delta function $\delta_{h_2 h_4}$ reminds that
the interference term is present only when the equal-flavour quarks are
indistinguishable, \emph{i.e.} have the same helicity.
The production rate for $q\qb\to q \qb \gamma\gamma$ has the same form as
\eqn{4quarkrate}, where we use \eqn{squamp4idquark}.
The production rates for $q q\to q q \gamma\gamma$, and for the
substitution of one or of both of the incoming quarks with antiquarks,
are obtained by suitable relabelings of the entries in 
\eqns{4quarkrate}{squamp4idquark}. However, note that when the
incoming fermions are both quarks or both antiquarks, the final-state
symmetry factor is 1/4 instead of 1/2.

When computing the virtual corrections we need the one-loop amplitude
for the $q\qb\to\gamma\gamma g$ process. Its colour decomposition is
similar to that of the tree-level amplitude given in \eqn{TwoQuarkGluonPhot},
\beq
\cA^\oneloop_5(1_\qb,2_q;3_g;4_\gamma,5_\gamma) = 2 e^2 \gs^3 
(T^{a_3})^{\ib_1}_{i_2}
A_{5:1}(1_\qb,2_q;3_g;4_\gamma,5_\gamma)\:,
\label{1loopcoldec}
\eeq 
where the one-loop colour sub-amplitude $A_{5:1}$ is given in terms of
sums of permutations of the primitive amplitudes for the
$q\qb\to g g g$ process~\cite{DelDuca:2000pa,Signer:1995nk}. Explicitly,
\bea
&&
A_{5;1}(1_\qb,2_q;3_g;4_\gamma,5_\gamma) =
  -Q_q^2\sum_{\sigma\in Z_2}
    A_{5}^{R,[1]}(1_\qb,3_g,2_q,{\sigma_4}_g,{\sigma_5}_g)
\nn\\ &&\qquad
-\sum_{\sigma\in S_3}
\left[{Q_q^2\over N_c^2}
     A_{5}^{R,[1]}(1_\qb,2_q,{\sigma_3}_g,{\sigma_4}_g,{\sigma_5}_g)
     -{\tr\left[Q^2_{f}\right]\over N_c}
A_5^{L,[{1\over2}]}(1_\qb,2_q,{\sigma_3}_g,{\sigma_4}_g,{\sigma_5}_g)
\right]\:,
\hspace*{2em}
\label{1looppart}
\eea
where $\tr\left[Q^2_{f}\right]$ is a short-hand for the sum of squared 
charges of different flavour appearing in the quark loop. We only need
the primitive amplitudes for the helicity configurations of 
type $(-,-,+,+,+)$\footnote{The amplitudes for the configurations with all
equal helicities, or all but one, as well as the amplitudes for
$g g\to g \gamma\gamma$ contribute at one loop, however they do not appear
in the NLO production rate since their tree level counterparts vanish.}.
The unrenormalized primitive amplitudes $A_5^{L,[{1\over2}]}$
can be found directly in Ref.~\cite{Bern:1995fz}. The $A_{5}^{R,[1]}$
must be first rewritten in terms of another set of primitive amplitudes (see 
Ref.~\cite{DelDuca:2000pa,Bern:1995fz})\footnote{In Ref.~\cite{Bern:1995fz}
the amplitudes $A_{5:1}$ are provided in the dimensional reduction 
scheme~\cite{Siegel:1979wq,Capper:1980ns,Bern:1992aq}. If it is wished to
have them in conventional dimensional regularization, one must add
to \eqn{1looppart} the term $A_5^\delta = - c_{\Gamma} (N_c^2-1)/(2N_c^2) 
A_5^\tree$, with $c_{\Gamma}$ in \eqn{cgam}.}.

The amplitude (\ref{1looppart}) is renormalized by adding the 
$\overline{\rm MS}$  ultraviolet counterterm,
\beq
\cA_5^\ct(1_\qb,2_q;3_g;4_\gamma,5_\gamma) = - c_{\Gamma} \gs^2 
{\beta_0\over 2\eps} \cA_5^\tree(1_\qb,2_q;3_g;4_\gamma,5_\gamma)\:.
\label{counter}
\eeq
with $\beta_0=(11N_c-2N_f)/3$, and 
\begin{equation}
c_{\Gamma} = {1\over 
(4\pi)^{2-\epsilon}}\, {\Gamma(1+\epsilon)\,
\Gamma^2(1-\epsilon)\over \Gamma(1-2\epsilon)}\:.
\label{cgam}
\end{equation}
The interference term between the one-loop and the tree amplitudes, summed
over colours, is
\bea
\lefteqn{
2 \sum_{\rm col} \re \left[ 
\cA^\oneloop_5(1_\qb,2_q;3_g;4_\gamma,5_\gamma) 
\cA_5^\tree(1_\qb,2_q;3_g;4_\gamma,5_\gamma)^* \right] } \nn\\ &=&
8 e^4 Q_q^2 \gs^4\,N_c^2 C_F\,\re \left[ 
A_{5:1}(1_\qb,2_q;3_g;4_\gamma,5_\gamma) 
A_5^\tree(1_\qb,2_q;3_g;4_\gamma,5_\gamma)^* \right]\:.
\label{interf}
\eea
Finally, the one-loop production rate for $q\qb\to g \gamma\gamma$ is given by
\beq
\d\sigma_v(q\qb) = {1\over 2\hat{s}}\ \d\cP_3\ {1\over 2}\ 
{1\over 4N_c^2}\ 2 \sum_{\rm col,\,hel} \re \left[ 
\cA^\oneloop_5(1_\qb,2_q;3_g;4_\gamma,5_\gamma) 
\cA_5^\tree(1_\qb,2_q;3_g;4_\gamma,5_\gamma)^* \right]\:.
\label{virtrate}
\eeq
The one-loop production rates for $q g \to q \gamma\gamma$ and 
$\qb g \to \qb \gamma\gamma$ are obtained by relabeling in \eqn{interf}
and by using the suitable colour average.

In addition to the NLO real and loop matrix elements, for the complete
NLO calculation we also need (i) a set of independent colour
projections of the matrix element squared at the Born level, summed
over parton polarizations, and (ii) an additional projection of the
Born level matrix element over the helicity of the external gluon in
four dimensions. These are required for the construction of the
subtraction terms~\cite{Catani:1996vz}.  The Born level process
involves three coloured partons. In this instance, the colour structure exactly
factorises, \emph{i.e.} the colour-correlated squared matrix elements
$|\cM_5^{i,k}|^2$ are proportional to the Born squared matrix element,
\bea
&&
|\cM_5^{i,k}(1_\qb,2_q;3_g;4_\gamma,5_\gamma)|^2 =
c_{ik} 
\sum_{\rm col,\,hel} |\cA_5^\tree(1_\qb,2_q;3_g;4_\gamma,5_\gamma)|^2\:,
\label{ccsquamp}
\eea
where $c_{ik} = c_{ki}$ and $c_{12} = 1/N_c$, $c_{13} = c_{23} = - N_c$.
The projection of the Born level matrix element over the helicity of
the external gluon is 
\bea
&&
\sum_{\rm col,\,hel}
\cA_5^\tree(1_\qb,2_q;3_g^+;4_\gamma,5_\gamma)
\cA_5^\tree(1_\qb,2_q;3_g^-;4_\gamma,5_\gamma)^* =
8 e^4 Q_q^4 \gs^2 N_c C_F
\nn\\ &&\qquad\qquad
\times \Big[
A_5^\tree(1_\qb^+,2_q^-;3_g^+;4_\gamma^-,5_\gamma^+)
A_5^\tree(1_\qb^+,2_q^-;3_g^-;4_\gamma^-,5_\gamma^+)^*
\nn\\ &&\qquad\qquad\quad
+ A_5^\tree(1_\qb^+,2_q^-;3_g^+;4_\gamma^+,5_\gamma^-)
A_5^\tree(1_\qb^+,2_q^-;3_g^-;4_\gamma^+,5_\gamma^-)^*
\Big]\:.
\eea

\subsection{Checks of the computations}
\label{sec:checks}

In order to find the NLO correction to the Born cross sections of
infrared-safe observables we used the dipole subtraction method
\cite{Catani:1996vz} with cut phase space for the subtraction term
as described in \cite{Nagy:1998bb}. The details of setting up the
partonic Monte Carlo programs based upon the cross section formulae
described in the previous subsections are well documented in
Refs.~\cite{Catani:1996vz} and \cite{Nagy:1998bb}. Here we record
the checks we performed to ensure the correctness of our prediction for
the NLO correction:
(i) we compared the Born cross section in \eqn{prodrate} to the
Born-level prediction of Ref.~\cite{deFlorian:1999tp} and found perfect
agreement;
(ii) we made three independent coding of the relevant NLO squared matrix
elements;
(iii) we checked numerically that the subtraction terms match the
singular behaviour of the real correction pointwise in randomly chosen
Monte Carlo points;
(iv) we found that the total NLO correction is independent, as it
should, of the parameter that controls the cut in the phase space of
the subtraction term (for details see Ref.~\cite{Nagy:1998bb});
(v) to utilize the process independent feature of the dipole method, we
used the same program library {\tt NLOJET++} that was used for other,
independently tested NLO computations, only the matrix elements were
changed.
  
\section{Results}
\label{sec:results}

We now turn to the presentation of our numerical predictions. We show
cross section values at leading order and at NLO for a Large Hadron
Collider (LHC) running at 14\,TeV. The values shown at
leading order were obtained using the leading order parton distribution
functions (p.d.f.'s) and those at NLO accuracy were obtained using the
NLO p.d.f.'s of the CTEQ6 package~\cite{Pumplin:2002vw} (tables cteq6l1
and cteq6m, respectively).  We used the two-loop running of the strong
coupling at NLO with $\Lambda_{\rm QCD} = 226$\,MeV and one-loop
running with $\Lambda_{\rm QCD} = 165$\,MeV at leading order. 
The renormalization and factorization scales were set to
$\mu_R = \mu_F = x_\mu \mu_0$, where for the reference value $\mu_0$
we used
\beq
\mu_0^2 = \frac14\left(\mgg^2 + \ptjet^2\right)\:, \label{eq:scale}
\eeq
with $\mgg$ the invariant mass of the photon pair. \eqn{eq:scale} 
reduces to the usual scale choice for inclusive photon pair production
if $\ptjet$ vanishes. Our prediction for the $\gamma\gamma\jet$
production cross section is intended for use in
the detection of a Higgs boson lighter than the top quark, therefore,
we assume 5 massless flavours and restrict all cross sections to the
range of 80\,GeV $\le \mgg \le$ 160\,GeV. The electromagnetic coupling
is taken at the Thomson limit, $\aem = 1/137$.

Our main goal in this paper is to show the general features of the
radiative corrections to the cross sections. Therefore, we use a
jet reconstruction algorithm and a set of event selection cuts,
expected to be typical in Higgs searches. In particular, in order to find the 
jet, we use the midpoint cone algorithm~\cite{Blazey:2000qt} with a cone 
size of $R = \sqrt{\Delta \eta^2 + \Delta \phi^2} = 1$, with $\Delta \eta$
the rapidity interval and $\Delta \phi$ the azimuthal angle\footnote{In 
our NLO computation the midpoint and seedless cone
algorithms yield identical cross sections.}.
Then, we require that both the jet and the photons have $\pt > 40\,$GeV
and rapidity within $|\eta| < 2.5$. These are the same selection cuts
as used in Ref.~\cite{deFlorian:1999tp} for computing the gluon
initiated $\ord(\as^3)$ corrections. Furthermore, we isolate both
photons from the partons in a cone of size $R_\gamma$.

At NLO the isolated photon cross section is not infrared safe. To
define an infrared safe cross section, one has to allow for some
hadronic activity inside the photon isolation cone. In a parton level
calculation it means that soft partons up to a predefined maximum
energy are allowed inside the isolation cone.

The standard way of defining an isolated prompt photon cross section,
that matches the usual experimental definition, is to allow for
transverse hadronic energy inside the photon isolation cone up to
$E_{\perp,{\rm max}} = \varepsilon p_{\gamma\perp}$, with
typical values of $\varepsilon$ between 0.1 and 0.5, and where
$p_{\gamma\perp}$ is taken either to be the photon transverse momentum
on an event-by-event basis or to correspond to the minimum value in the
$p_{\gamma\perp}$ range. In perturbation
theory this isolation requires the splitting of the cross section into
a direct and a fragmentation contribution. The precise definition of
the two terms, valid to all orders in perturbation theory, is given in
Ref.~\cite{Catani:2002ny}, where it was shown that a small isolation
cone for the photon leads to unphysical results in a fixed order
computation.  For a small cone radius $R_\gamma$, an all-order resummation of
$\as \ln(1/R_\gamma^2)$ terms combined with a careful study of the
border line between perturbative and non-perturbative regions has to be
undertaken. To avoid such problems, we use $R_\gamma = 0.4$ as the
default radius.

In this work, we deal with only the direct production of two photons,
therefore, we use a `smooth' photon-isolation prescription which
depends on the production of direct photons only and is independent of the
fragmentation contribution~\cite{Frixione:1998jh}. This isolation means
that the energy of the soft parton inside the isolation cone has to
converge to zero smoothly if the distance in the $\eta-\phi$ plane
between the photon and parton vanishes. Explicitly, the amount of hadronic
transverse energy  $E_\perp$ (which in our NLO partonic computation is
equal to the transverse momentum of the possible single parton in the
isolation cone) in all cones of radius $r < R_\gamma$ must be less than
\beq
E_{\perp,{\rm max}} = \varepsilon p_{\gamma\perp}
\left(\frac{1 - \cos r}{1 - \cos R_\gamma}\right)^n\:,
\label{eqn:frixione}
\eeq
where we chose to use $n = 1$ and $\varepsilon = 0.5$ as default values,
and we take $p_{\gamma\perp}$ to be the photon transverse momentum
on an event-by-event basis.
\EPSFIGURE[t]{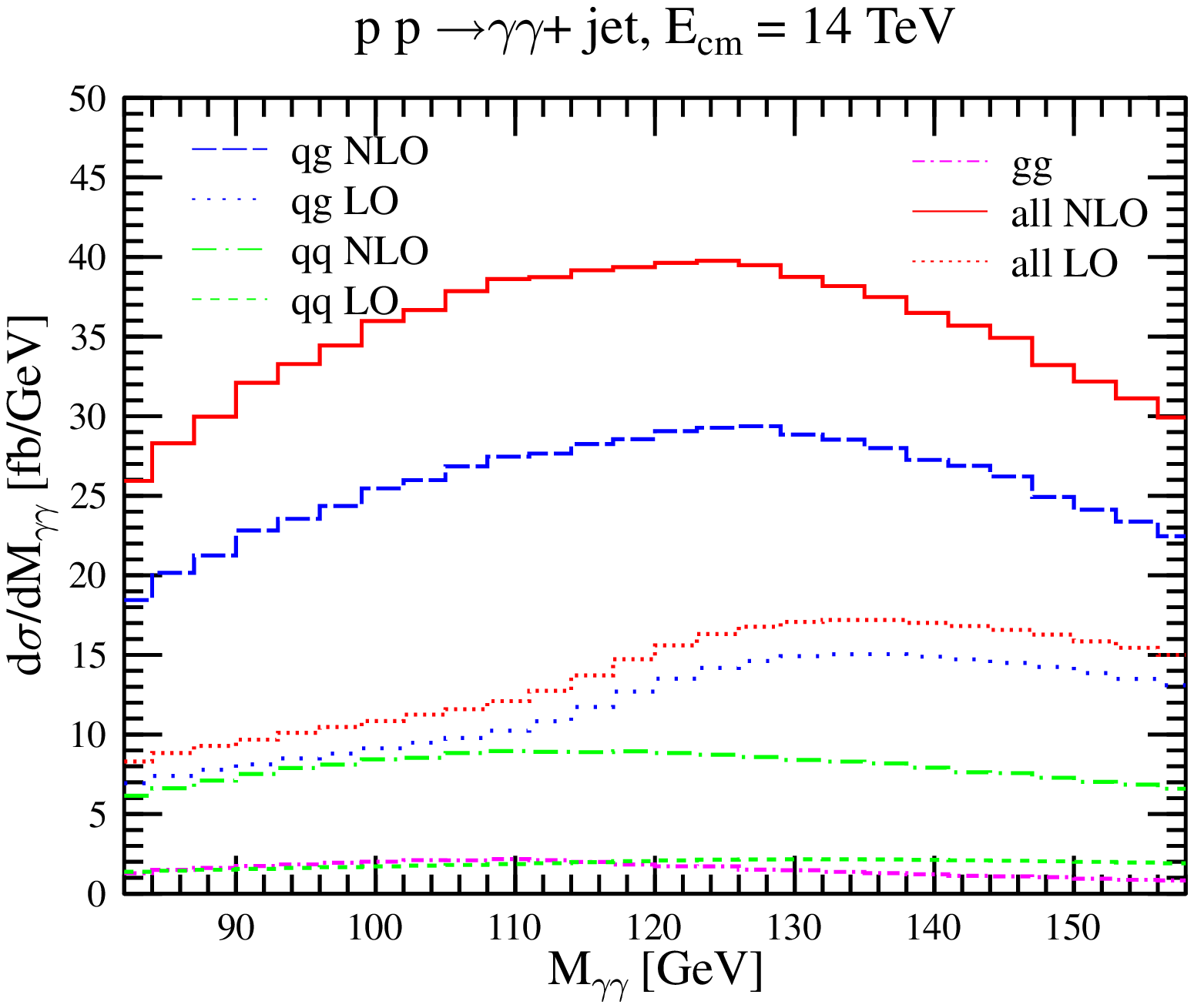,width=.9\textwidth,clip=}
{The invariant mass distribution of the photon pair at LHC energy.
The photons and the jet are required to have transverse momentum
$|p_\perp| \ge 40$~GeV and lie in the central rapidity region of
$|\eta| \le 2.5$. The jet is reconstructed according to the
midpoint algorithm.
\label{fig:mgg} }

In Fig.\,\ref{fig:mgg} we plot the invariant mass distribution of the
photon pair. Here we see the continuum background on which the Higgs
signal is expected to manifest itself as a narrow resonance in the
intermediate-mass range. The dotted (red) line is the leading order
prediction and the solid (red) one is the differential cross section at
NLO accuracy.  The striking feature of the plot is the rather large
correction.  The large corrections are partly due to the appearance of
new subprocesses at NLO as can also be read off the figure. The
gluon-gluon scattering subprocess begins to appear only at NLO accuracy,
and therefore it is effectively leading order. It is shown with a long
dashed-dotted (magenta) line: it yields a very small contribution. 
The bulk of the cross section comes from quark-gluon scattering both at
leading order and at NLO, shown with sparsely-dotted (blue) and
long-dashed (blue) lines. The quark-quark scattering receives rather
large corrections because the leading
order subprocess can only be a quark-antiquark annihilation process,
shown with short-dashed (green) line, while at NLO, shown with short
dashed-dotted (green) line, there are (anti-)quark-(anti-)quark
scattering subprocesses. Thus at NLO the parton luminosity is sizeably
larger. In addition, more dynamic processes are allowed, which include
$t$-channel gluon exchange. These contribute to enlarge the cross section
in phase space regions which are disfavoured at leading order.

A part of the large radiative corrections is accounted for by the new
subprocesses; another part is due simply to the enlarged phase space,
as can be seen from Fig.~\ref{fig:rjga}, where the differential
distributions in the distance
$R_{j\gamma} = \sqrt{|\eta_j-\eta_\gamma|^2 + |\phi_j-\phi_\gamma|^2}$
between the jet and the photons in the $\eta$-$\phi$ plane are shown,
with a selection cut at $R_{j\gamma} \ge 0.4$. We have also studied the
distribution of the rapidity separation between the harder photon and the
jet and we found that it is on average small. Therefore, the peak in the
$R_{j\gamma_h}$ distribution near $\pi$ suggests that the distribution 
peaks where the jet and the harder photon are mostly back-to-back and
central in rapidity.  The harder photon is separated from the jet by at
least $R_{j\gamma_h} \approx 2$ at leading order, while at NLO they can
be as close as permitted by the selection cut.  On the contrary, in
going from leading order to NLO, the distribution for the photon with
lower transverse momentum, the \emph{softer} photon, changes in
normalization but not in shape.
\EPSFIGURE[t]
{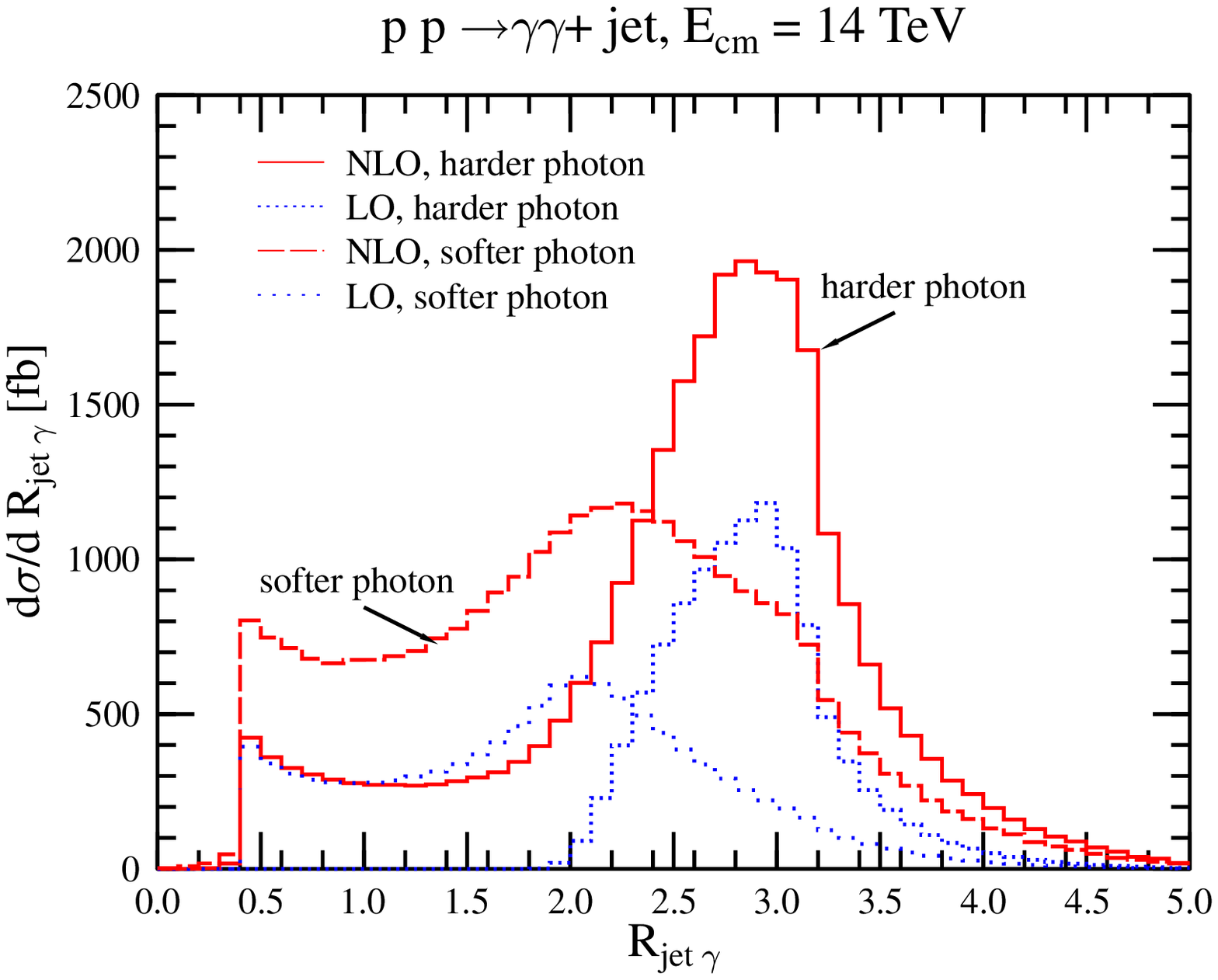,width=.9\textwidth}
{The distributions in the distances between the jet and the photons in
the $\eta$-$\phi$ plane. The same selection cuts are used as in Fig.~1.
\label{fig:rjga} }

In the following plots, we require $R_{j\gamma} \ge 1.5$, which is also
a standard selection cut in Higgs searches in the Higgs + jet
channel~\cite{cms,atlas}. In fact, on one hand a cut on
$R_{j\gamma_s} > 1.5$ affects the leading order and the NLO evaluation
in the same way because they have the same shape. Thus in that case the
cut does not reduce the correction to the $\mgg$ distribution. On the
other hand, a cut on $R_{j\gamma_h} > 1.5$ cuts the NLO correction, but
does not cut the leading order, so the correction is reduced.
Nevertheless, the reduction is less then 10\,\%, thus cutting on
$R_{j\gamma}$ at $R_{j\gamma} \ge 1.5$ reduces the NLO correction to
the invariant mass distribution of the photon pair, but only marginally.
\EPSFIGURE[t]
{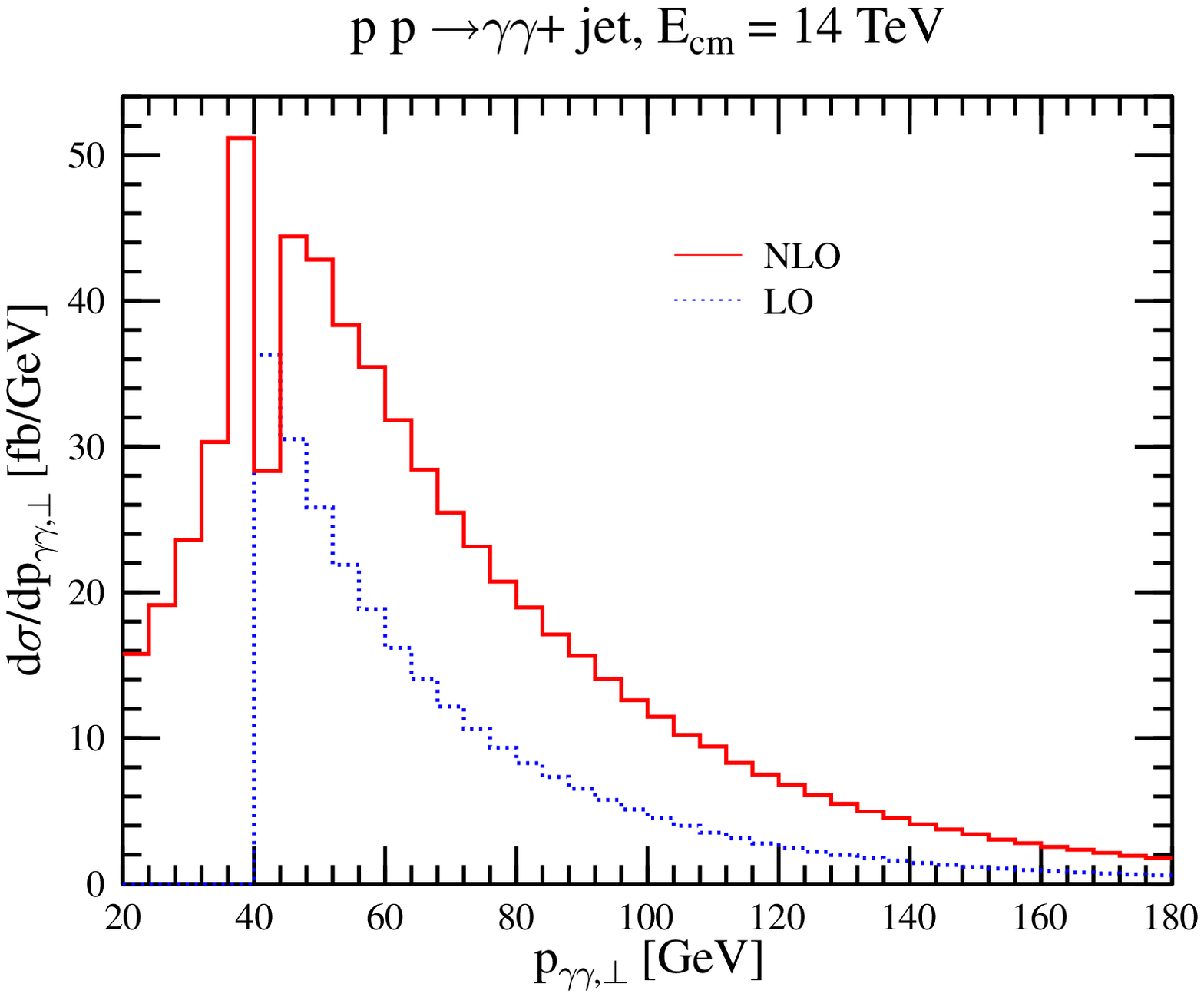,width=.9\textwidth}
{Transverse momentum distribution of the photon pair.
The same selection cuts are used as in Fig.~1.
\label{fig:ptgaga} }

In Fig.~\ref{fig:ptgaga}, we plot the differential distribution of the
transverse momentum of the photon pair,
$\ptgg = |\bom{p}_{\gamma_1,\perp} + \bom{p}_{\gamma_2,\perp}|$, with a
cut at $p_{{\rm jet} \perp}^{\rm min} = 40$~GeV.  At leading order the
jet recoils against the photon pair and the respective jet and photon
pair $p_\perp$ distributions are identical.  At NLO the extra parton
radiation opens the
part of the phase space with $\ptgg < p_{{\rm jet} \perp}^{\rm min}$.
The double peak around 40\,GeV is an artifact of the fixed-order
computation, similar to the NLO prediction at $C = 0.75$ for the
$C$-parameter distribution in electron-positron annihilation.  The
fixed-order calculation is known to be unreliable in the vicinity of
the threshold, where an all-order resummation is necessary
\cite{Catani:1997xc}, which would result in a structure, called Sudakov 
shoulder, that is
continuous and smooth at $\ptgg = p_{{\rm jet} \perp}^{\rm min}$. 
Without the resummation, we must introduce a cut, $p_{\gamma\gamma,\perp}
\ge 40$\,GeV to avoid regions in the phase space where the fixed-order
prediction is not reliable. In the following plots, in addition to the
default cuts, we also require $p_{\gamma\gamma,\perp} \ge 40$\,GeV.

The magnitude of the NLO corrections is also heavily influenced by the
photon-isolation cut parameters. In Figs.~\ref{fig:mgg1}--\ref{fig:mgg3}
we show the effect of scanning the parameters $R_\gamma$ and $\eps$ in
\eqn{eqn:frixione} over the values of $R_\gamma = 0.4$, 1 and
$\eps = 0.1$, 0.5, 1. We see that the leading order predictions are 
independent of these parameters\footnote{The leading-order cross section
does not depend on the chosen values of $R_\gamma$ because we have
required $R_{j \gamma} \ge 1.5$ and the jet momentum is the same as that
of the only parton in the final state.}, but the NLO ones depend
strongly on the photon isolation. Firstly, we note that the smaller
$R_\gamma$ the larger the NLO corrections. In addition, the larger
$\eps$ the larger the NLO correction, with the effect being larger if
$R_\gamma$ is larger.  This behaviour is in agreement with \eqn{eqn:frixione},
according to which smaller $R_\gamma$ and larger $\epsilon$ imply
larger amounts of soft-parton energy that is allowed inside the cone.
\DOUBLEFIGURE[t]
{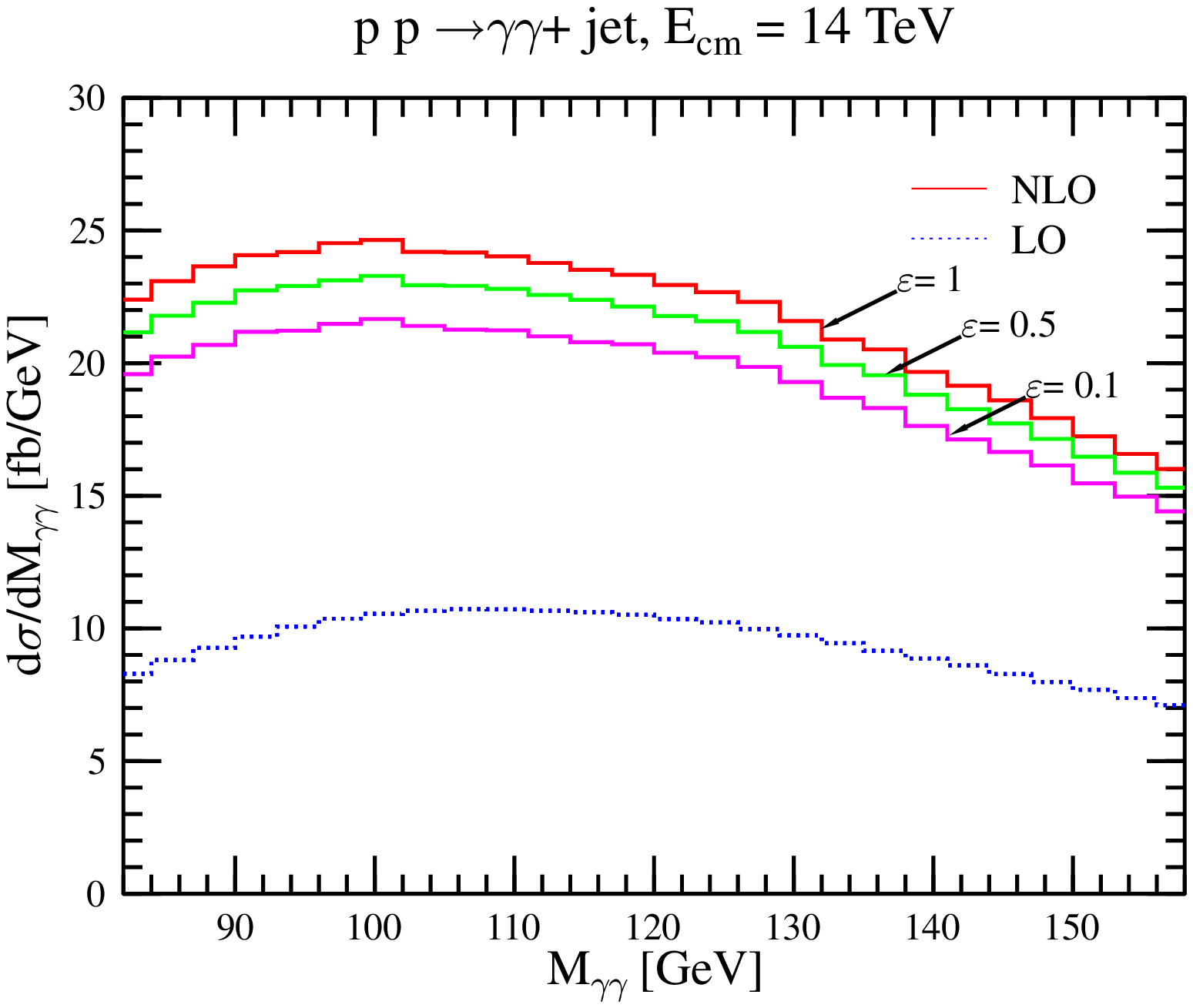,width=.45\textwidth}
{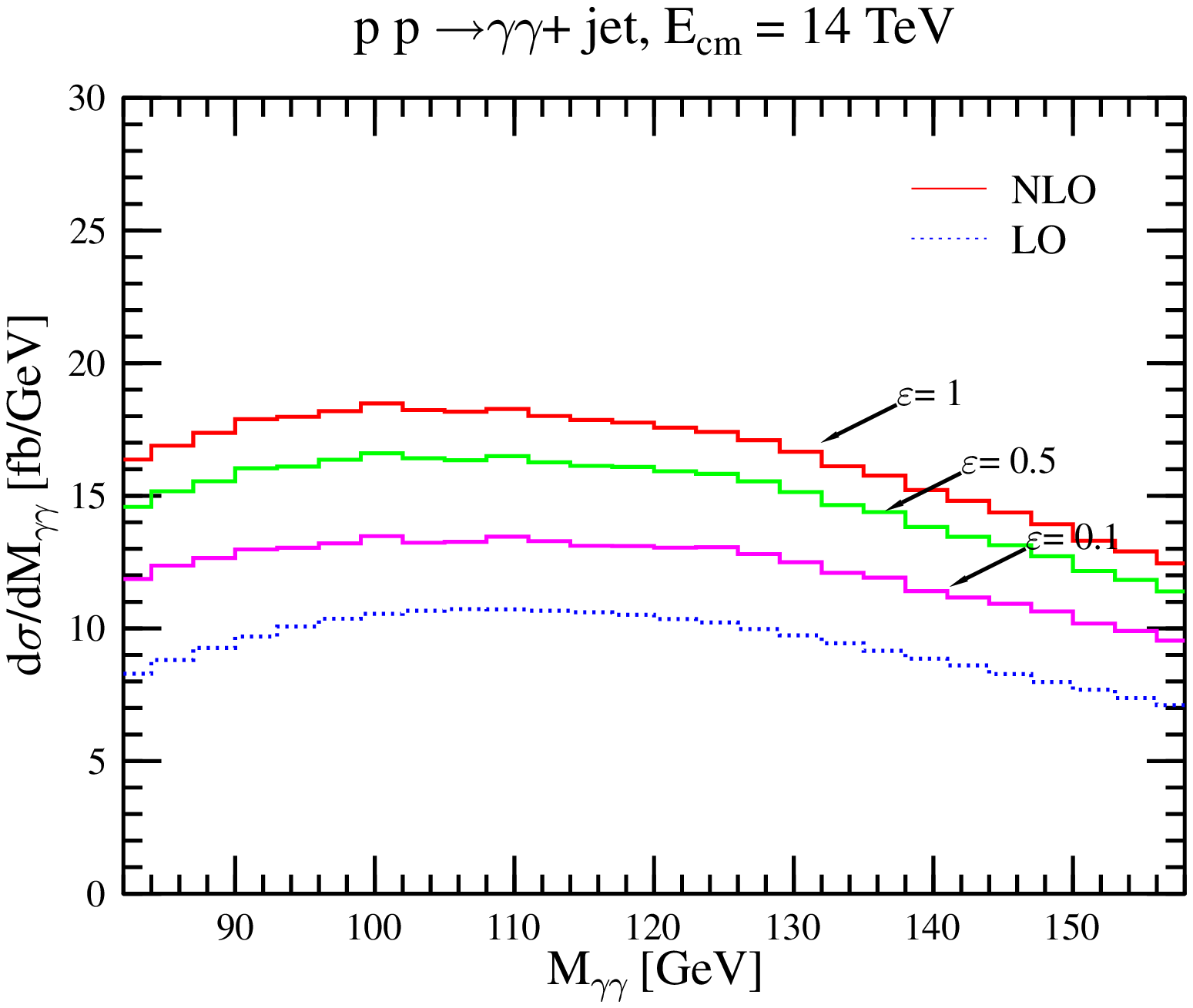,width=.45\textwidth}
{Dependence of the invariant mass distribution of the photon pair on the
photon isolation parameter $\eps$ for $R_\gamma = 0.4$.
In addition to the selection cuts used in Fig.~1, we also
require $R_{j\gamma} \ge 1.5$ and $p_{\gamma\gamma,\perp} \ge 40$\,GeV.
\label{fig:mgg1} }
{Dependence of the invariant mass distribution of the photon pair on the
photon isolation parameter $\eps$ for $R_\gamma = 1$.
The same selection cuts are used as in Fig.~4.
\label{fig:mgg3} }

Another remarkable feature of Figs.~\ref{fig:mgg1}--\ref{fig:mgg3} is
that with the applied cuts, the two-photon plus jet background for the
search of a Higgs boson with mass in the 120--140\,GeV range is rather
flat, therefore, well measurable from the sidebands around the
hypothetical Higgs signal. This feature is very different from the shape
of the background to the inclusive $pp\to H \to \gamma\gamma$ channel,
which is steeply falling.

\EPSFIGURE[t]
{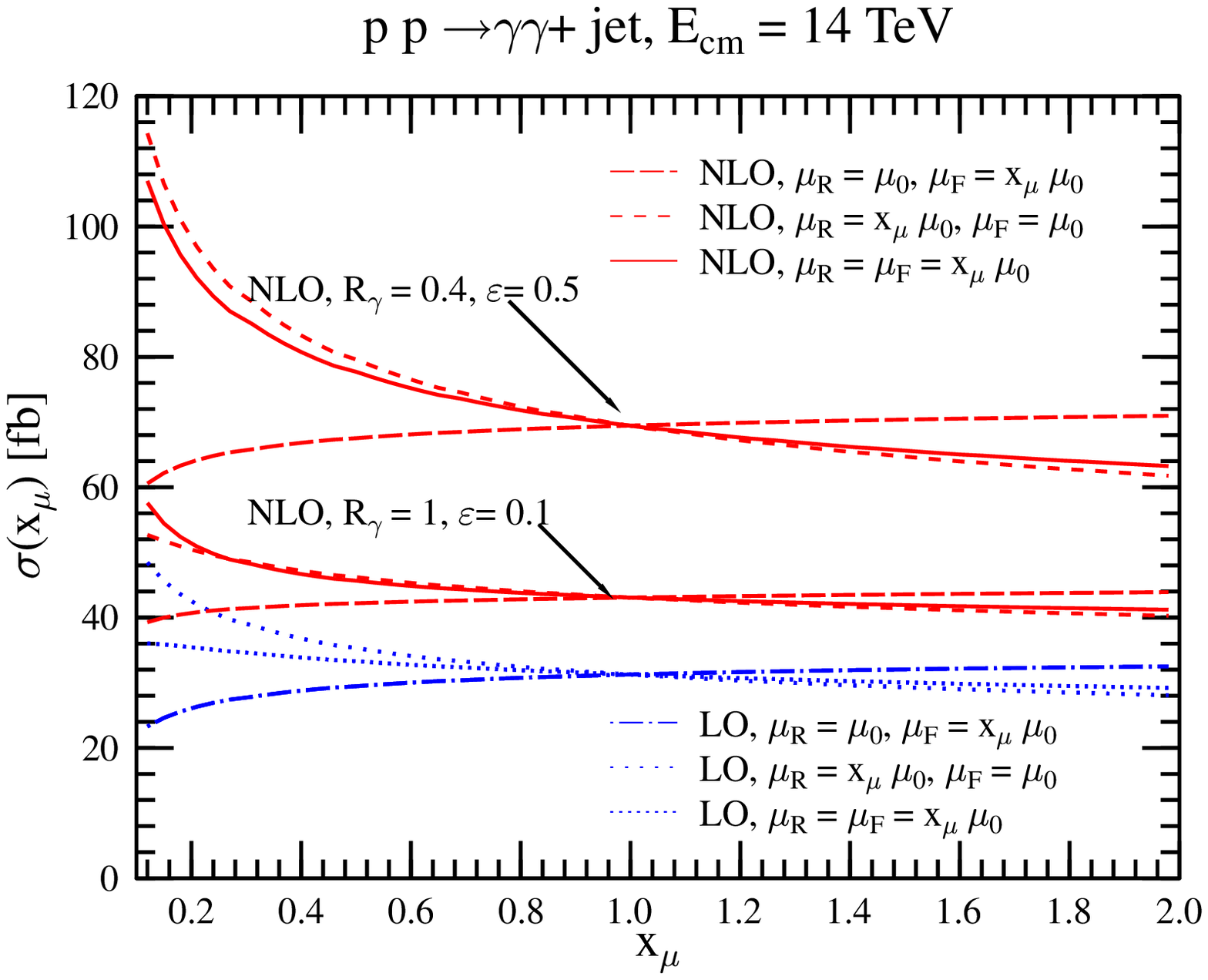,width=.9\textwidth}
{Dependence on the renormalization and factorization scales of the cross
section in a bin of 118.5\,GeV $\le \mgg \le$ 121.5\,GeV at leading-order
and at NLO accuracy. The same selection cuts are used as in Fig.~4.
\label{fig:xmu} }
In order to assess the stability of the predictions against scale
variations, we show the cross section in a 3\,GeV bin around $\mgg =
120$\,GeV, that is the background for a hypothetical Higgs signal for
a Higgs particle of mass 120\,GeV. \fig{fig:xmu} shows the cross section
for two sets of photon-isolation parameters. We show the scale variations
for varying the renormalization and factorization scales separately,
keeping the other scale fixed, as well as varying them simultaneously.
The lower three curves represent the leading order predictions.  At
leading order the dependence on the renormalization and factorization
scales is rather small, especially when the two scales are set equal
(densely dotted
line). Observing the predictions we conclude that the scale dependence at
leading order does not represent the uncertainty of the predictions due
to the unknown higher orders. The inclusion of the radiative corrections 
results mainly in the substantial increase of the cross section.
For our default isolation ($R_\gamma = 0.4$ and $\epsilon = 0.5$)
the NLO corrections at $x_\mu = 1$ are about 120\,\% of the leading
order prediction.  In addition, the scale dependence is not reduced by
the inclusion of the radiative corrections. If we require more
stringent photon isolation cuts, then we find smaller corrections and a
more stable prediction.  For instance, in \fig{fig:xmu} we also show
the scale dependence of the cross section obtained with $R_\gamma = 1$
and $\eps = 0.1$. We find that the cross section at NLO is
about 40\,\% larger than the leading order prediction, and in this case
the scale dependence at NLO is reduced as compared to the one at
leading order accuracy.  

\section{Conclusions}
\label{sec:conclusions}

In this paper we computed the QCD radiative corrections to the
$p p\to\gamma\gamma\jet$ process that constitutes part of
the irreducible background to the $p p\to H\jet \to \gamma\gamma\jet$
discovery channel of an intermediate-mass Higgs boson at the LHC.
We used a smooth photon isolation that is infrared safe to all orders in
perturbation theory and independent of the photon fragmentation into
hadrons. The predictions were made with a partonic Monte Carlo program
that can be utilized for a detailed study of the signal significance
for the intermediate-mass Higgs boson discovery in the Higgs + jet
production channel at the LHC.  We found large radiative corrections,
however they are rather sensitive to the selection cuts and photon
isolation parameters.  Choosing a mild photon isolation, \emph{i.e.}\ a
small isolation cone radius $R_\gamma = 0.4$ with relatively large
hadronic activity allowed in the cone results in more than 100\,\%
correction with as large residual scale dependence at NLO as at leading
order. Making the photon isolation more stringent, for instance
increasing the cone radius to $R_\gamma = 1$ and decreasing the
hadronic activity in the cone reduces both the magnitude of the
radiative corrections as well as the dependence on the renormalization
and factorization scales. This result shows that a constant $K^{\rm NLO}
= 1.6$ factor, as mentioned in the Introduction, is certainly not
appropriate for taking into account the radiative corrections to the
irreducible background of the $pp \to H\jet \to \gamma\gamma\jet$
discovery channel at the LHC.  

\section*{Acknowledgments}
We thank S.~Catani and S.~Frixione for useful discussions.
ZN and ZT thank the INFN, sez. di Torino, VDD thanks the
Nucl.~Res.~Inst.~of the HAS for their kind hospitality during the late
stage of this work.
This work was supported in part by the EU Fourth Framework Programme
``Training and Mobility of Researchers'', Network ``QCD and particle
structure'', contract FMRX-CT98-0194 (DG 12 - MIHT)
and by the Hungarian Scientific Research Fund grants OTKA T-038240.


\end{document}